\documentclass[twocolumn,showpacs,superscriptaddress,preprintnumbers,amsmath,amssymb,prb]{revtex4}

\usepackage{graphicx}
\usepackage{dcolumn}
\usepackage{bm}
\usepackage{amssymb}
\usepackage{color,soul}
\usepackage{float}

\begin{document}

\title{Aharonov-Bohm effect for excitons in a semiconductor quantum ring dressed by circularly polarized light}

\author{O. V. Kibis}\email{Oleg.Kibis@nstu.ru}
\affiliation{Department of Applied and Theoretical Physics,
Novosibirsk State Technical University, Karl Marx Avenue 20,
Novosibirsk 630073, Russia} \affiliation{Division of Physics and
Applied Physics, Nanyang Technological University 637371,
Singapore}

\author{H. Sigurdsson}
\affiliation{Division of Physics and Applied Physics, Nanyang
Technological University 637371, Singapore}

\author{I. A. Shelykh}
\affiliation{Division of Physics and Applied Physics, Nanyang
Technological University 637371, Singapore} \affiliation{Science
Institute, University of Iceland, Dunhagi-3, IS-107, Reykjavik,
Iceland}\affiliation{ITMO University, St. Petersburg 197101,
Russia}

%\date{\today}

\begin{abstract}
We show theoretically that the strong coupling of circularly
polarized photons to an exciton in ring-like semiconductor
nanostructures results in physical nonequivalence of clockwise and
counterclockwise exciton rotations in the ring. As a consequence,
the stationary energy splitting of exciton states corresponding to
these mutually opposite rotations appears. This excitonic
Aharonov-Bohm effect depends on the intensity and frequency of the
circularly polarized field and can be detected in state-of-the-art
optical experiments.
\end{abstract}


\maketitle

\section{Introduction} Progress in semiconductor
nanotechnologies has led to developments in the fabrication of
various mesoscopic objects, including quantum rings. The
fundamental physical interest attracted by these systems arises
from a wide variety of purely quantum-mechanical effects which can
be observed in ring-like nanostructures. One of them is the
Aharonov-Bohm (AB) effect arisen from the direct influence of the
vector potential on the phase of the electron wave function.
\cite{AharonovBohm, Chambers1960} In ring-like nanostructures
pierced by a magnetic flux, this effect results in the energy
splitting of electron states corresponding to mutually opposite
directions of electronic rotation in the ring.
\cite{Buttiker_1983} As a consequence, magnetic-flux-dependent
oscillations of the conductance of the ring appear.
\cite{Gefen1983,Buttiker1984,Webb,Timp,Wees,Shelykh2005} Since the
AB effect takes place for both a single electron and many-particle
quantum states, \cite{Sutherland_1990} it can be observed for
elementary excitations in semiconductor nanostructures as well.
The simplest of them is an exciton which is a bound quantum state
of a negative charged electron in the conduction band and a
positive charged hole in the valence band. Manifestations of
various excitonic effects in semiconductor ring-like structures,
including the AB effect induced by a magnetic field, have
attracted great attention of both theorists
\cite{Chaplik_1995,Romer_2000,Song_2001,Hu_2001,Govorov_2002,Silva_2004,Moulopoulos_2004,Kovalev_2007,Willatzen_2009,Bagheri_2010,Santander_2011,Li_2011}
and experimentalists.
\cite{Lorke_2000,Bayer_2003,Ribeiro_2004,Haft_2002,Ding_2010}

Fundamentally, the AB effect is caused by the broken time-reversal
symmetry in an electron system  subjected to a magnetic flux.
Namely, the flux breaks the equivalence of clockwise and
counterclockwise electron rotation inside a ring-like structure,
which results in the flux-controlled interference of the electron
waves corresponding to these rotations. The similar broken
equivalence of electron motion for mutually opposite directions
caused by a magnetic field can take place in various
nanostructures, including quantum wells, \cite{Kibis_1997} carbon
nanotubes \cite{Romanov_1993} and hybrid semiconductor/ferromagnet
nanostructures. \cite{Kibis2002_1} However, the time-reversal
symmetry can be broken not only by a magnetic flux but also by a
circularly polarized electromagnetic field. Indeed, the field
breaks the symmetry since the time reversal turns clockwise
polarized photons into counterclockwise polarized ones and vice
versa. In quantum rings, the strong electron coupling to
circularly polarized photons results in the magnetic-flux-like
splitting of electron energy levels corresponding to mutually
opposite electronic rotation in the ring \cite{Kibis2011} and
oscillations of the ring conductance as a function of the
intensity and frequency of the irradiation. \cite{Sigurdsson_2014}
This phenomenon can be described in terms of a stationary
artificial $U(1)$ gauge field generated by the strong coupling
between an electron and circularly polarized photons.
\cite{Sigurdsson_2014} Therefore, various stationary phenomena
similar to the AB effect can take place in ring-like electronic
systems interacting with a circularly polarized electromagnetic
field. As a consequence, the new class of quantum optical
phenomena in semiconductor nanostructures appears. Although a
theory of these AB-like phenomena in quantum rings has been
elaborated for a single electron, \cite{Kibis2011,Sigurdsson_2014}
the optically-induced AB effect for excitons still awaits detailed
analysis. The given article is aimed to fill this gap in the
theory, which lies at the border between quantum optics and
physics of semiconductor nanostructures.

The paper is organized as follows. In Section II, the
exciton-photon Hamiltonian is analyzed and solutions of the
exciton-photon Schr\"odinger problem are found. In Section III,
the energy spectrum of the dressed excitons is discussed and
experimental sets to detect the effect are proposed. In Section
IV, conclusion is presented.

\section{Model} 
\begin{figure}
\centering
\includegraphics[width=0.40\textwidth]{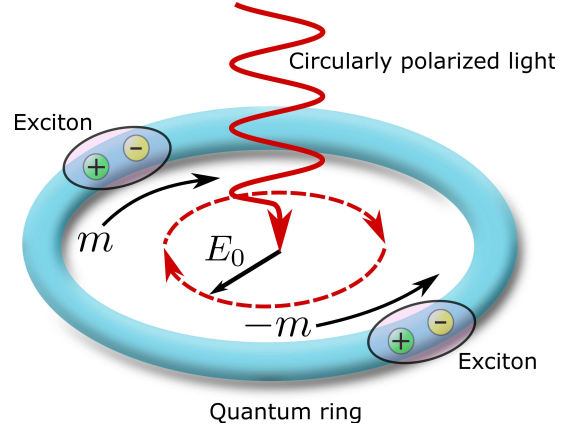}
\caption{(Color online) Sketch of an exciton-field system in a
quantum ring under consideration. The exciton coupling to the
circularly polarized electromagnetic field results in physical
nonequivalence of exciton states corresponding to clockwise and
counterclockwise rotations of the exciton as a whole along the
ring (shown by the arrows). These exciton states are described by
mutually opposite angular momenta $m$ and $-m$ along the ring
axis.} \label{fig.ring}
\end{figure}

An electron-hole pair in an one-dimensional
quantum ring (see Fig.~\ref{fig.ring}) can be described by the
Hamiltonian
\begin{equation}\label{H}
\hat{\cal{H}}_0=-\frac{\hbar^2}{2m_hR^2}\frac{\partial^2}{\partial\varphi_h^2}
-\frac{\hbar^2}{2m_eR^2}\frac{\partial^2}{\partial\varphi_e^2}
+V(\varphi_e-\varphi_h),
\end{equation}
where $R$ is the radius of the ring, $m_{e,h}$ is the effective
masse of an electron (hole) in the ring, $V(\varphi_e-\varphi_h)$
is the potential energy of hole-electron interaction, and
$\varphi_{e,h}$ is the azimuthal angle of the electron (hole) in
the ring. Introducing the new variables,
$$\varphi=\frac{m_e\varphi_e+m_h\varphi_h}{m_e+m_h}, \qquad \theta=\varphi_e-\varphi_h,$$
the Hamiltonian (\ref{H}) can be rewritten as
\begin{equation}\label{H1}
\hat{\cal{H}}_0=-\frac{\hbar^2}{2MR^2}\frac{\partial^2}{\partial\varphi^2}
-\frac{\hbar^2}{2\mu R^2}\frac{\partial^2}{\partial\theta^2}
+V(\theta),
\end{equation}
where $M=m_e+m_h$ is the exciton mass, and $\mu=m_em_h/M$ is the
reduced exciton mass. The eigenfunctions of the stationary
Schr\"odinger equation with the Hamiltonian (\ref{H1}) have the
form
\begin{equation}\label{psi}
\psi_{nm}(\varphi,\theta)=\chi_n(\theta)\frac{e^{im\varphi}}{\sqrt{2\pi}},
\end{equation}
where the function $\chi(\theta)$ meets the Schr\"odinger equation
\begin{equation}\label{shr}
-\frac{\hbar^2}{2\mu
R^2}\frac{\partial^2\chi_{n}(\theta)}{\partial\theta^2}+V(\theta)\chi_{n}(\theta)=\varepsilon_n\chi_n(\theta),
\end{equation}
$m=0,\pm1,\pm2,...$ is the exciton angular momentum along the ring
axis, $n=0,1,2,...$ is the principal quantum number of the
exciton, and $\varepsilon_n$ is the exciton binding energy.
Correspondingly, the full energy of exciton reads as
\begin{equation}\label{Ex}
\varepsilon_{n,m}=\varepsilon_n+\frac{\hbar^2m^2}{2MR^2},
\end{equation}
where the second term is the kinetic energy of rotational motion
of exciton in the ring.

Let the ring be subjected to a circularly polarized
electromagnetic wave with the frequency $\omega$, which propagates
along the ring axis (see Fig.~\ref{fig.ring}). Then the full
Hamiltonian of the exciton-photon system, including both the field
energy, $\hbar\omega\hat{a}^\dagger\hat{a}$, and the exciton
Hamiltonian, $\hat{\mathcal H}_0$, is
\begin{equation}\label{H0}
\hat{\mathcal
H}=\hbar\omega\hat{a}^\dagger\hat{a}+\hat{\cal{H}}_0+\hat{U},
\end{equation}
where $\hat{a}$ and $\hat{a}^\dagger$ are the operators of photon
annihilation and creation, respectively, written in the
Schr\"{o}dinger representation (the representation of occupation
numbers), and $\hat{U}$ is the operator of exciton-photon
interaction. Generalizing the operator of electron-photon
interaction in a quantum ring \cite{Kibis2011} for the considered
case of electron-hole pair, we can write this operator as
\begin{equation}\label{UR}
\hat{U}=\frac{iq_eR}{2}\sqrt{\frac{\hbar\omega}{\epsilon_0V_0}}
\left[(e^{-i\varphi_e}-e^{-i\varphi_h})\hat{a}^\dagger+(e^{i\varphi_h}
-e^{i\varphi_e}) \hat{a}\right],
\end{equation}
where $q_e$ is the electron charge, $V_0$ is the quantization
volume, and $\epsilon_0$ is the vacuum permittivity. To describe
the exciton-photon system, let us use the notation $|n,m,N\rangle$
which indicates that the electromagnetic field is in a quantum
state with the photon occupation number $N=1,2,3,...\;$, and the
exciton is in a quantum state with the wave function (\ref{psi}).
The electron-photon states $|n,m,N\rangle$ are true eigenstates of
the Hamiltonian $$\hat{\mathcal
H}^{(0)}_{R}=\hbar\omega\hat{a}^\dagger\hat{a}+\hat{\cal{H}}_0,$$
which describes the non-interacting exciton-photon system.
Correspondingly, their energy spectrum is
$$\varepsilon^{(0)}_{n,m,N}=N\hbar\omega+\varepsilon_{n,m}.$$ In
order to find the energy spectrum of the full electron-photon
Hamiltonian (\ref{H0}), let us use the conventional perturbation
theory, considering the term (\ref{UR}) as a perturbation with the
matrix elements $\langle
n^\prime,m^\prime,N^\prime|\hat{U}|n,m,N\rangle$. Taking into
account in Eq.~(\ref{UR}) that $\varphi_e=\varphi+m_h\theta/M$ and
$\varphi_h=\varphi-m_e\theta/M$, these matrix elements read as
\begin{eqnarray}
\langle
n^\prime,m^\prime,N^\prime|\hat{U}|n,m,N\rangle=eR\sqrt{\frac{\hbar\omega}{\epsilon_0V_0}}
\nonumber\\
\times\left[I_{n^\prime
n}\sqrt{N+1}\delta_{m,m^\prime+1}\delta_{N,N^\prime-1}\right.\nonumber\\
\left.-I_{n^\prime
n}^\ast\sqrt{N}\delta_{m,m^\prime-1}\delta_{N,N^\prime+1}\right],
\end{eqnarray}
where $$I_{n^\prime
n}=\int_{-\pi}^{\pi}\chi^\ast_{n^\prime}(\theta)\chi_{n}(\theta)e^{-i(m_h-m_e)\theta/2M}\sin(\theta/2)\mathrm{d}\theta.$$
Performing trivial calculations within the second order of the
perturbation theory, we can derive eigenenergies of the
exciton-photon Hamiltonian (\ref{H0}),
\begin{eqnarray}\label{ES}
\varepsilon_{n,m,N}&=&\varepsilon^{(0)}_{n,m,N}\nonumber\\
&+&\sum_{n^\prime}\left[\frac{|\langle
n^\prime,m+1,N-1|\hat{U}|n,m,N\rangle|^2}{\varepsilon^{(0)}_{n,m,N}-\varepsilon^{(0)}_{n^\prime,m+1,N-1}}\right.\nonumber\\
&+&\left.\frac{|\langle n^\prime,
m-1,N+1|\hat{U}|n,m,N\rangle|^2}{\varepsilon^{(0)}_{n,m,N}-\varepsilon^{(0)}_{n^\prime,m-1,N+1}}\right].
\end{eqnarray}
Since Eq.~(\ref{ES}) is derived within the second order of the
perturbation theory, it describes the problem correctly if the
energy differences in denominators of all terms lie far from zero.
In what follows, we have to keep in mind that all parameters of
the problem must lie far from these resonant points.

The energy spectrum of exciton-photon system (\ref{ES}) can be
written formally as
$\varepsilon_{n,m,N}=N\hbar\omega+\widetilde{\varepsilon}_{n,m,N}$,
where the first term is the field energy. Following the
conventional terminology of quantum optics,
\cite{QuantumOptics1,QuantumOptics2} the second term,
$\widetilde{\varepsilon}_{n,m,N}$, is the energy spectrum of the
exciton dressed by the circularly polarized field (dressing
field). Restricting our analysis by the most interesting case of
classically strong dressing field ($N\gg1$), we arrive from
Eq.~(\ref{ES}) to the sought energy spectrum of dressed exciton,
\begin{eqnarray}\label{Eex}
\widetilde{\varepsilon}_{n,m}&=&\sum_{n^\prime}\left[\frac{(q_eE_0R)^2|I_{nn^\prime}|^2}
{\varepsilon_{n,m}-\varepsilon_{n^\prime,m+1}+\hbar\omega}\right.\nonumber\\
&+&\left.\frac{(q_eE_0R)^2|I_{nn^\prime}|^2}
{\varepsilon_{n,m}-\varepsilon_{n^\prime,m-1}-\hbar\omega}\right],
\end{eqnarray}
where $E_0=\sqrt{N\hbar\omega/\epsilon_0V_0}$ is the classical
amplitude of electric field of the electromagnetic wave. It is
apparent that dressed exciton states with mutually opposite
angular momenta, $m$ and $-m$, have different energies
(\ref{Eex}). Physically, this should be treated as a field-induced
nonequivalence of clockwise and counterclockwise exciton rotations
in the ring. As a consequence, the excitonic Aharonov-Bohm effect
induced by the circularly polarized field appears. In order to
simplify the calculation of the field-induced splitting,
$\Delta\widetilde{\varepsilon}_{n,m}=\widetilde{\varepsilon}_{n,m}-\widetilde{\varepsilon}_{n,-m}$,
we will restrict our consideration to the case of the ground
exciton state with $n=0$. Let us assume that the characteristic
binding energy of exciton, $q_e^2/4\pi\epsilon_0R^2$, is much more
than both the characteristic energy of rotational exciton motion,
$\hbar^2|m|/2MR^2$, and the photon energy $\hbar\omega$. Then we
can neglect the field-induced mixing of exciton states with
$n^\prime\neq0$ in Eq.~(\ref{Eex}). As a result, we arrive from
Eq.~(\ref{Eex}) to the field-induced splitting of exciton states
with mutually opposite angular momenta,
\begin{eqnarray}\label{split}
\Delta\widetilde{\varepsilon}_{0,m}=
\left|\int_{-\pi}^{\pi}|\chi_{0}(\theta)|^2\sin\left(\frac{m_h-m_e}{2M}\,\theta\right)\sin\left(\frac{\theta}{2}\right)\mathrm{d}\theta\right|^2\nonumber\\
\times\left[\frac{2\hbar\omega(q_eE_0R)^2}
{\varepsilon_R^2(1-2m)^2-(\hbar\omega)^2}
-\frac{2\hbar\omega(q_eE_0R)^2}
{\varepsilon_R^2(1+2m)^2-(\hbar\omega)^2}\right],\nonumber\\
\end{eqnarray}
where $\varepsilon_R=\hbar^2/2MR^2$ is the characteristic energy
of exciton rotation. In order to calculate the integral in
Eq.~(\ref{split}), we have to solve the Schr\"odinger equation
(\ref{shr}) and find the wave function $\chi_{0}(\theta)$.
Approximating the electron-hole interaction potential $V(\theta)$
in Eq.~(\ref{shr}) by the delta-function \cite{Romer_2000} and
assuming the characteristic exciton size,
$a=\hbar/\sqrt{8\mu\varepsilon_0}$, to be much less than the ring
length $2\pi R$, we can write the splitting (\ref{split}) in the
final form
\begin{eqnarray} \label{eq.sqbr}
\Delta\widetilde{\varepsilon}_{0,m}=\frac{\hbar\omega}{2}
\left(\frac{m_h-m_e}{M}\right)^2(eE_0a)^2\nonumber\\
\times\left[\frac{1} {\varepsilon_R^2(1-2m)^2-(\hbar\omega)^2}
-\frac{1}{\varepsilon_R^2(1+2m)^2-(\hbar\omega)^2}\right].
\end{eqnarray}
It should be stressed that the simplest delta-function model
\cite{Romer_2000} leads to reasonable results. This follows
formally from the fact that the final expression (\ref{eq.sqbr})
does not depend on model parameters: It depends only on the
exciton binding energy $\varepsilon_0$ which should be treated as
a phenomenological parameter. We checked that numerical
calculation using the Coulomb potential gives very similar results
to those obtained analytically for the case of the delta potential
if the binding energy $\varepsilon_0$ is kept the same.

Let us estimate the main limitation of the model one-dimensional
Hamiltonian (\ref{H}) which neglects the exciton motion in the
radial direction. It can be important since the radial motion
weakens the AB effect in wide rings. \cite{Santander_2011} Let a
ring with the radius $R$ has the width $\Delta R$. It follows from
the numerical calculations that amplitudes of the AB oscillations
for the case of $R/\Delta R>5$ and for the case of ideal
one-dimensional ring ($\Delta R\rightarrow0$) are almost
identical.\cite{Santander_2011} Therefore, the one-dimensional
Hamiltonian (\ref{H}) correctly describes the solved AB problem
for typical semiconductor rings with radius $R$ in the tens of
nanometers and width $\Delta R$ in the nanometer range.

\section{Results and discussion} The field-induced splitting
(\ref{split})--(\ref{eq.sqbr}) vanishes if the electron mass is
equal to the hole mass, $m_e=m_h$. Physically, this can be
explained in terms of an artificial $U(1)$ gauge field produced by
the coupling of a charged particle to circularly polarized
photons. \cite{Sigurdsson_2014} Since the artificial field
\cite{Sigurdsson_2014} depends on a particle mass, it interacts
differently with an electron and a hole in the case of $m_e\neq
m_h$. As a consequence, the splitting
(\ref{split})--(\ref{eq.sqbr}) is nonzero in the case of $m_e\neq
m_h$, though an exciton is electrically neutral as a whole. In the
case of $m_e=m_h$, the artificial gauge field interacts equally
with both electron and hole. However, signs of the interaction are
different for the electron and the hole since electrical charges
of electron and hole are opposite. Therefore, the interaction of
the artificial gauge field with an exciton is zero in the case of
$m_e=m_h$.

\begin{figure}
\centering
\includegraphics[width=0.50\textwidth]{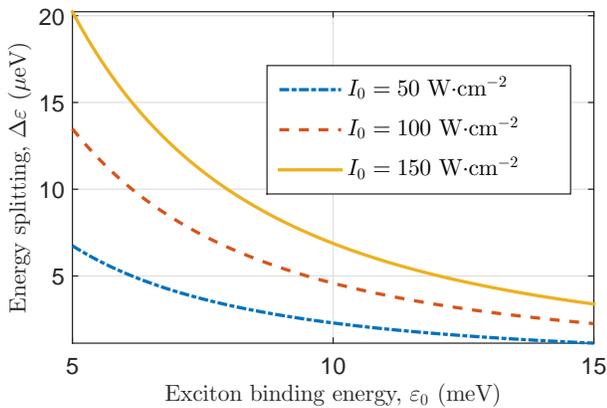}
\caption{(Color online) The energy splitting of the exciton states
with angular momenta $m=1$ and $m=-1$ in a GaAs ring with the
radius $R=9.6$ nm as a function of the exciton binding energy
$\varepsilon_0$ for a circularly polarized dressing field with the
frequency $\omega=1050$ GHz and different intensities $I_0$.}
\label{fig2}
\end{figure}
\begin{figure}
\centering
\includegraphics[width=0.52\textwidth]{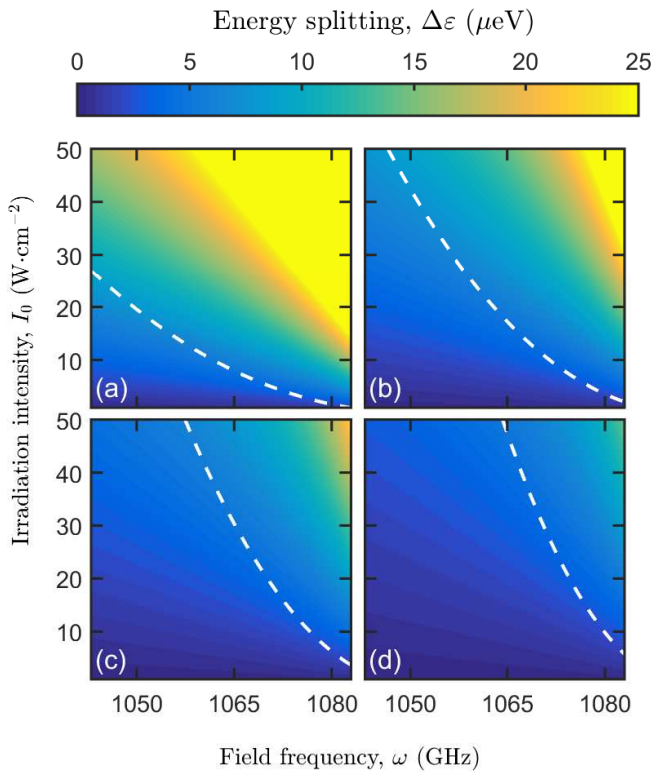}
\caption{(Color online) The energy splitting of exciton states
with angular momenta $m=1$ and $m=-1$ in a GaAs ring with the
radius $R=9.6$ nm as a function of the field intensity $I_0$ and
the field frequency $\omega$ for different binding energies of the
exciton: (a) $\varepsilon_0=2$ meV; (b) $\varepsilon_0=4$ meV; (c)
$\varepsilon_0=6$ meV; (d) $\varepsilon_0=8$ meV. The physically
relevant areas of the field parameters, which correspond to
applicability of the basic expressions derived within the
perturbation theory, lie below of the dashed lines.} \label{fig3}
\end{figure}
The splitting (\ref{eq.sqbr}) for exciton states with the angular
momenta $m=1$ and $m=-1$ in a GaAs quantum ring is presented
graphically in Figs.~(\ref{fig2})--(\ref{fig3}) for various
intensities of the dressing field, $I_0=\epsilon_0E_0^2c$. The
used effective masses of electron and holes in GaAs, $m_e/m_0 =
0.063$ and $m_h/m_0 = 0.51$, are taken from
Ref.~\onlinecite{Handbook}, where $m_0$ is the mass of electron in
vacuum. In Fig.~(\ref{fig2}), the splitting
$\Delta\varepsilon=\widetilde{\varepsilon}_{0,1}-\widetilde{\varepsilon}_{0,-1}$
is plotted as a function of the exciton binding energy,
$\varepsilon_0$, which depends on the confinement potential of a
quantum ring. \cite{Song_2001} It is apparent that the splitting
decreases with increasing the binding energy. Physically, this is
a consequence of decreasing the exciton size, $a$. Indeed, an
exciton with a very small size looks like an electrically neutral
particle from viewpoint of the dressing electromagnetic field. As
a consequence, the splitting (\ref{eq.sqbr}) is small for small
excitons.

It follows from Figs.~(\ref{fig2})--(\ref{fig3}) that the typical
splitting is of $\mu$eV scale for stationary irradiation
intensities of tens W/cm$^2$. This splitting is comparable to the
Lamb shift in atoms and can be detected experimentally by optical
methods. It order to increase the splitting, the irradiation
intensity $I_0$ should also be increased. However, the increasing
of stationary irradiation can fluidize a semiconductor ring. To
avoid the fluidizing, it is reasonable to use narrow pulses of a
strong dressing field which splits exciton states and narrow
pulses of a weak probing field which detects the splitting. This
well-known pump-and-probe methodology is elaborated long ago and
commonly used to observe quantum optics effects --- particularly,
modifications of energy spectrum of dressed electrons arisen from
the optical Stark effect
--- in semiconductor structures (see, e.g., Refs.~\onlinecite{Joffre_1988_1,Joffre_1988_2,Lee_1991}). Within this approach,
giant dressing fields (up to GW/cm$^2$) can be applied to
semiconductor structures. As a consequence, the splitting
(\ref{eq.sqbr}) can be of meV scale in state-of-the-art optical
experiments.

It should be stressed that the discussed effect is qualitatively
different to those arisen from absorption of circularly polarized
light in quantum rings (see, e.g.,
Refs.~\onlinecite{Matos_2005,Pershin_2005,Nobusada_2005}). Namely,
the absorption of photons with non-zero angular momentum by
electrons leads to the transfer of angular momentum from light to
electrons in a ring. Correspondingly, photoinduced currents in the
ring appear.\cite{Matos_2005,Pershin_2005,Nobusada_2005} Since
this effect is caused by light absorption, it can be described
within the classical electrodynamics of ring-shaped conductors. In
contrast, we consider the Aharonov-Bohm effect induced by light in
the regime of electromagnetic dressing, when absorption of real
photons is absent. To be more specific, the discussed AB effect
arises from light-induced changing phase of electron wave
function, which results in the appearance of the artificial gauge
field \cite{Sigurdsson_2014} and shifts exciton energy levels in
the ring. Evidently, this purely quantum phenomenon cannot be
described within classical physics.

\section{Conclusion} Summarizing the aforesaid, we predict
a new quantum-optical phenomenon in semiconductor ring-like
nanostructures. Namely, a high-frequency circularly polarized
electromagnetic wave splits the energy levels of excitons in a
semiconductor quantum ring. This effect should be treated as an
optically-induced Aharonov-Bohm effect for excitons and can be
observed in quantum rings with using modern experimental technics.
It should be noted that, besides semiconductor quantum rings,
perspective objects for observing the discussed effect are such
ring-like semiconductor structures as carbon nanotubes.

\begin{acknowledgements} The work was partially supported
by FP7 IRSES projects POLATER and QOCaN, FP7 ITN project NOTEDEV,
Rannis project BOFEHYSS, RFBR project 14-02-00033, and the Russian
Ministry of Education and Science.
\end{acknowledgements}

\end{document}